\documentclass[fleqn,usenatbib]{mnras}

\usepackage{newtxtext,newtxmath}

\usepackage[T1]{fontenc}

\DeclareRobustCommand{\VAN}[3]{#2}
\let\VANthebibliography\thebibliography
\def\thebibliography{\DeclareRobustCommand{\VAN}[3]{##3}\VANthebibliography}

\usepackage{graphicx}	
\usepackage{amsmath}	

\newcommand{\Msun}{\ensuremath{\mathrm{M}_\odot}}

\newcommand{\tzo}{T{\.Z}O}
\newcommand{\tzos}{T{\.Z}Os}

\graphicspath{{./}{figures/}}

\title[\tzo\ explosion properties]{
Properties of Thorne-\.Zytkow object explosions
}

\author[T. J. Moriya and S. I. Blinnikov]{
Takashi J. Moriya$^{1,2}$\thanks{E-mail: takashi.moriya@nao.ac.jp (TJM)}
and Sergei I. Blinnikov$^{3,4,5}$
\\
$^{1}$National Astronomical Observatory of Japan, National Institutes of Natural Sciences, 2-21-1 Osawa, Mitaka, Tokyo 181-8588, Japan \\
$^{2}$School of Physics and Astronomy, Faculty of Science, Monash University, Clayton, Victoria 3800, Australia \\
$^{3}$ National Research Center "Kurchatov institute", Institute for Theoretical and Experimental Physics (ITEP), Moscow 117218, Russia \\
$^{4}$ Sternberg Astronomical Institute, Moscow State University, Moscow 119234, Russia \\
$^{5}$ Kavli Institute for the Physics and Mathematics of the Universe (WPI), The University of Tokyo Institutes for Advanced Study, The University of Tokyo, \\ 5-1-5 Kashiwanoha, Kashiwa, Chiba 277-8583, Japan 
}

\date{Accepted 2021 September 7. Received 2021 September 7; in original form 2021 July 26}

\pubyear{2021}

\begin{document}
\label{firstpage}
\pagerange{\pageref{firstpage}--\pageref{lastpage}}
\maketitle

\begin{abstract}
Thorne-\.Zytkow objects are stars that have a neutron star core with an extended hydrogen-rich envelope. Massive Thorne-\.Zytkow objects are proposed to explode when the nuclear reactions sustaining their structure are terminated by the exhaustion of the seed elements. In this paper, we investigate the observational properties of the possible Thorne-\.Zytkow object explosions. We find that Thorne-\.Zytkow object explosions are observed as long-duration transients lasting for several years. If the accretion disk triggering the explosions does not last for a long time, Thorne-\.Zytkow object explosions have a luminosity plateau with about $10^{39}~\mathrm{erg~s^{-1}}$ lasting for a few years, and then they suddenly become faint. They would be observed as vanished stars after a bright phase lasting for a few years. If the accretion disk is sustained for long time, the Thorne-\.Zytkow object explosions become as bright as supernovae. They would be observed as supernovae with rise times of several hundred days. We found that their photospheric velocities are $2000~\mathrm{km~s^{-1}}$ at most, much smaller than those found in supernovae. Supernovae with extremely long rise times such as HSC16aayt and SN~2008iy may be related to the explosions of Thorne-\.Zytkow objects.
\end{abstract}




\section{Introduction}
Thorne-\.Zytkow objects (\tzos) are hypothetical stars that have a neutron star core with an extended hydrogen-rich envelope \citep[][]{thorne1975tzo,thorne1977tzo}. Such stars can be formed through the spiral-in of a neutron star into a companion star following unstable mass transfer \citep[][]{taam1978tzo}, a supernova (SN) kick \citep[][]{leonard1994tzo} or an unsuccessful SN explosion \citep[][]{utrobin2008}.

Recent observations start to identify \tzo\ candidates. \citet{levesque2014tzo} identified a chemical anomaly in a supergiant star HV2112 in Small Magellanic Cloud \citep[][]{worley2016tzo,mcmillan2018} which is consistent with a \tzo\ \citep[][]{tout2014tzo}. It is also suggested that HV2112 may be an asymptotic-giant-branch star rather than a \tzo\ \citep[][]{tout2014tzo,beasor2018,ogrady2020}. There are ongoing observational efforts to discover \tzos\ \citep[][]{demarchi2021}.

Massive (more than about $16~\Msun$) \tzos, which are supported by the nuclear reactions near the neutron star surface \citep[][]{cannon1993tzo,biehle1991tzo}, are expected to collapse when they burn all the seed elements required for the nuclear reactions \citep[][]{Bisnovatyi-Kogan1984,podsiadlowski1995tzofate}. \citet{moriya2018} suggested that massive \tzo\ collapses may lead to their explosion. In this study, we investigate the outcomes of the potential \tzo\ explosions and show their expected observational properties.

The rest of this paper is organized as follows. First, we show our assumptions and methods to investigate the \tzo\ explosion properties in Section~\ref{sec:method}. Then we present the expected \tzo\ explosion properties in Section~\ref{sec:results}. We discuss the observational properties and possible \tzo\ explosion candidates in Section~\ref{sec:discussion}. We summarize this paper in Section~\ref{sec:summary}.

\section{Methods}\label{sec:method}
In this section, we briefly introduce our \tzo\ explosion model and our method to estimate the properties of \tzo\ explosions.

\subsection{Progenitor and explosion trigger}
\begin{figure}
	\includegraphics[width=\columnwidth]{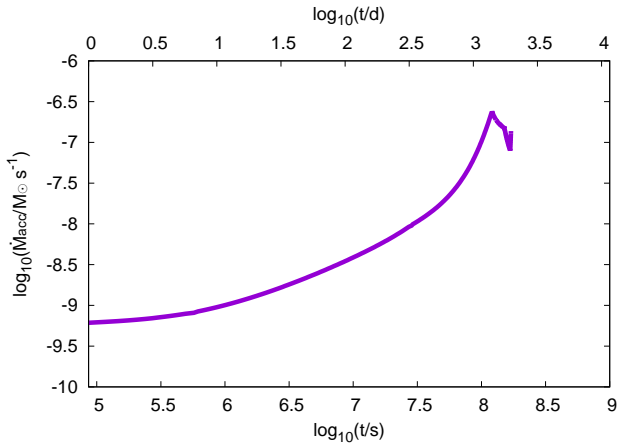}
    \caption{
    Accretion rate $\dot{M}_\mathrm{acc}$ towards the central compact object after accretion disk formation that is estimated to occur at around 1~day after the onset of the \tzo\ collapse. The time $t$ is after the onset of the \tzo\ collapse.
    }
    \label{fig:accretionrate}
\end{figure}

We take the 16~\Msun\ \tzo\ model of \citet{biehle1991tzo} as we did in our previous study \citep{moriya2018}. Assuming a typical \tzo\ angular velocity,
\citet{moriya2018} showed that the accretion disk around the central neutron star is expected to appear at around 1~day after the onset of collapse when $10^{-3}~\Msun$ is accreted. The subsequent accretion time-scale becomes about 10 times longer than the free-fall time-scale, if we assume a typical viscosity parameter of 0.1. The estimated accretion rate $\dot{M}_\mathrm{acc}$ after 1~d since the onset of the collapse is presented in Fig.~\ref{fig:accretionrate}.

\begin{figure}
	\includegraphics[width=\columnwidth]{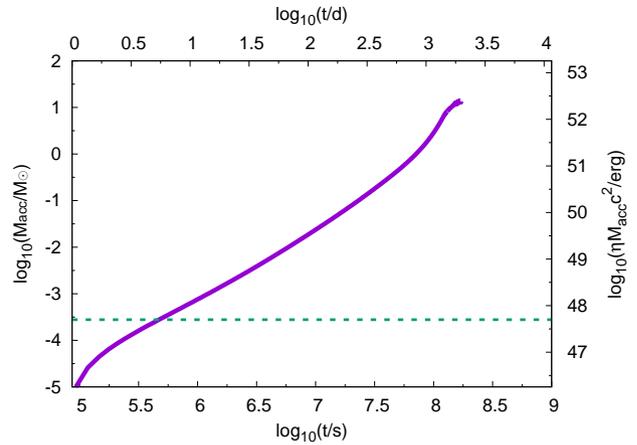}
    \caption{Mass accreted on to the central compact object after accretion disk formation. The time $t$ is after the onset of the \tzo\ collapse. The right axis shows the injected energy to the collapsing \tzo\ envelope through the accretion with an efficiency $\eta = 10^{-3}$. The dashed line shows the binding energy ($5\times 10^{47}~\mathrm{erg}$) of the \tzo\ modelled.}
    \label{fig:accretionmass}
\end{figure}

Once the accretion disk is formed, a fraction ($\eta$) of accreted energy could be released at the central region of the collapsing \tzo\ through an accretion disk wind or jet. When the energy sufficient to unbind the \tzo\ has been released an explosion of \tzo\ can be triggered. The energy input rate $\dot{E}_\mathrm{in}$ at the centre is expressed as
\begin{equation}
    \dot{E}_\mathrm{in} = \eta \dot{M}_\mathrm{acc} c^2, \label{eq:Ein}
\end{equation}
where $c$ is the speed of light. We adopt an efficiency $\eta = 10^{-3}$ in this study \citep[e.g.,][]{dexter2013kasen}. Fig.~\ref{fig:accretionmass} shows the total accreted mass $M_\mathrm{acc}$ after the disk formation at 1~day after the collapse and the total injected energy $\eta M_\mathrm{acc}c^2$. In this work, we mainly consider the cases where the central energy injection occurs at wide angles, not in collimated jets. 

Once the energy injection from the accretion disk has released more energy than the binding energy of the \tzo, the \tzo\ can explode. However, the accretion energy injection could be terminated by the outflows that push back the accreting materials. It is not clear when the accretion is terminated and so we introduce another parameter, $t_\mathrm{acc}$, the time for which the accretion continues. Because the binding energy of the 16~\Msun\ \tzo\ is $5\times 10^{47}~\mathrm{erg}$, 
the accretion needs to last at least for about 5~d to cause the \tzo\ to explode (Fig.~\ref{fig:accretionmass}). The \tzo\ explosion energy depends on $t_\mathrm{acc}$. For $t_\mathrm{acc}=10,100$ and 1000~d the explosion energies are $1.1\times 10^{48}~\mathrm{erg}$, $3.5\times 10^{49}~\mathrm{erg}$ and $3.3\times 10^{51}~\mathrm{erg}$, respectively.

\subsection{Radiation hydrodynamics calculations}
In order to estimate the observational properties of \tzo\ explosions, we use the one-dimensional multi-frequency radiation hydrodynamics code \texttt{STELLA}. We refer to \citet[][]{blinnikov1998sn1993j,blinnikov2000sn1987a,blinnikov2006sniadeflg} for the full details of the code. Briefly, \texttt{STELLA} calculates time-dependent equations of hydrodynamics and the angular moments of intensity averaged over a frequency bin with the variable Eddington method \citep[][]{mihalas1984}. Spectral energy distributions (SEDs) are numerically evaluated at every time-step. When we define a photosphere, we take the location where the Rosseland-mean optical depth becomes $2/3$. In this work, we put thermal energy with the rate of Eq.~\ref{eq:Ein} at 1~\Msun\ from the centre. This is the surface of the central neutron star supporting the \tzo. The energy injection begins 1~d after the onset of the collapse and continues until $t_\mathrm{acc}$. Though we use a one-dimensional code to investigate the \tzo\ explosion properties, the energy injection from the accretion disk is not necessarily spherically symmetric. The \tzo\ explosion properties that we show here are applicable when the energy injection occurs on a large angular scale. Our models are not applicable if the energy injection occurs on a small angular scale through such as in collimated jets.

\section{Results}\label{sec:results}
\begin{figure}
	\includegraphics[width=\columnwidth]{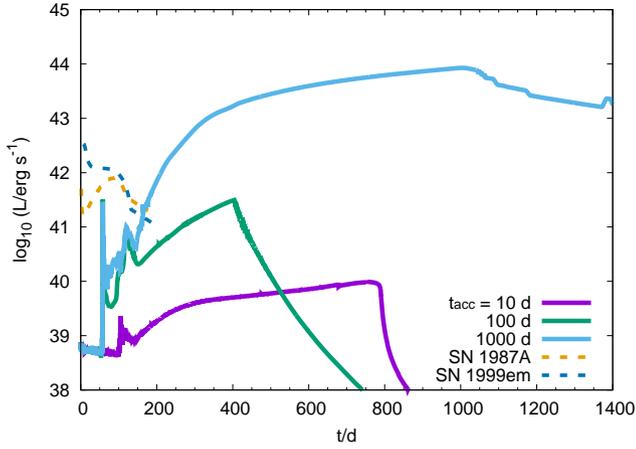}
    \caption{Synthetic bolometric luminosity $L$ of the \tzo\ explosions with $t_\mathrm{acc}=10, 100$ and 1000~d. The time $t$ is after the onset of the \tzo\ collapse. The bolometric light curves of SN~1987A \citep{hamuy1988} and SN~1999em \citep{bersten2009} are shown for comparison.}
    \label{fig:bolometric}
\end{figure}

\begin{figure}
	\includegraphics[width=\columnwidth]{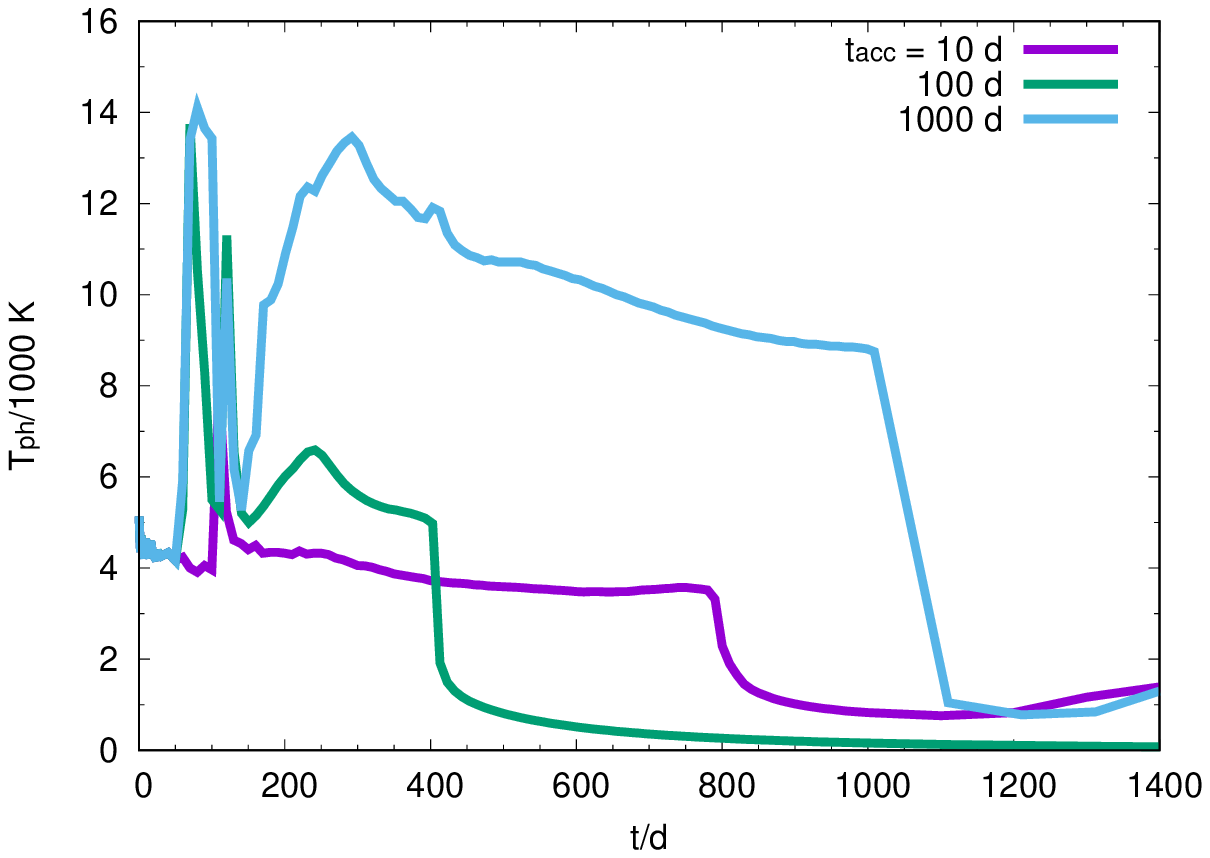}
	\includegraphics[width=\columnwidth]{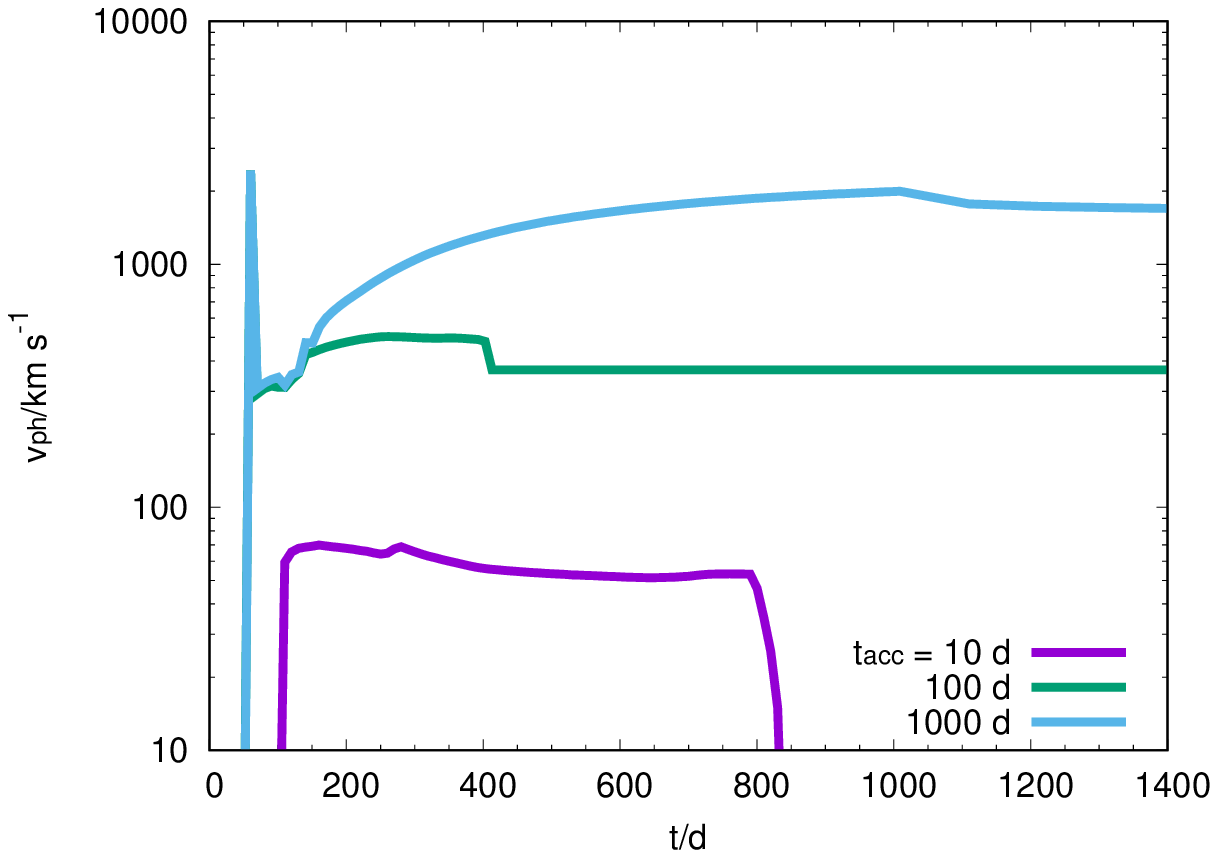}
    \caption{Photospheric temperature $T_\mathrm{ph}$ (top) and velocity $v_\mathrm{ph}$ (bottom) evolution of the \tzo\ explosions. The time $t$ is after the onset of the \tzo\ collapse.}
    \label{fig:photosphere}
\end{figure}

\begin{figure}
	\includegraphics[width=\columnwidth]{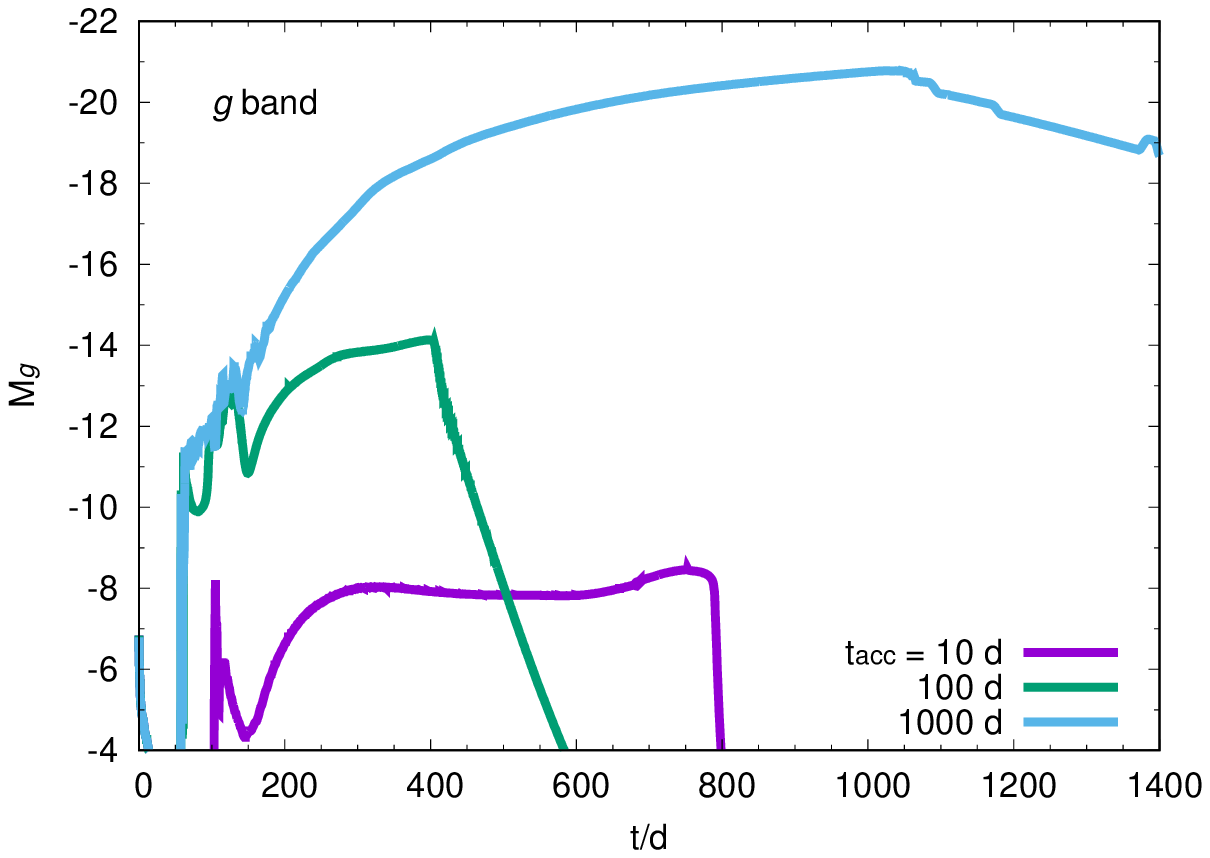}
	\includegraphics[width=\columnwidth]{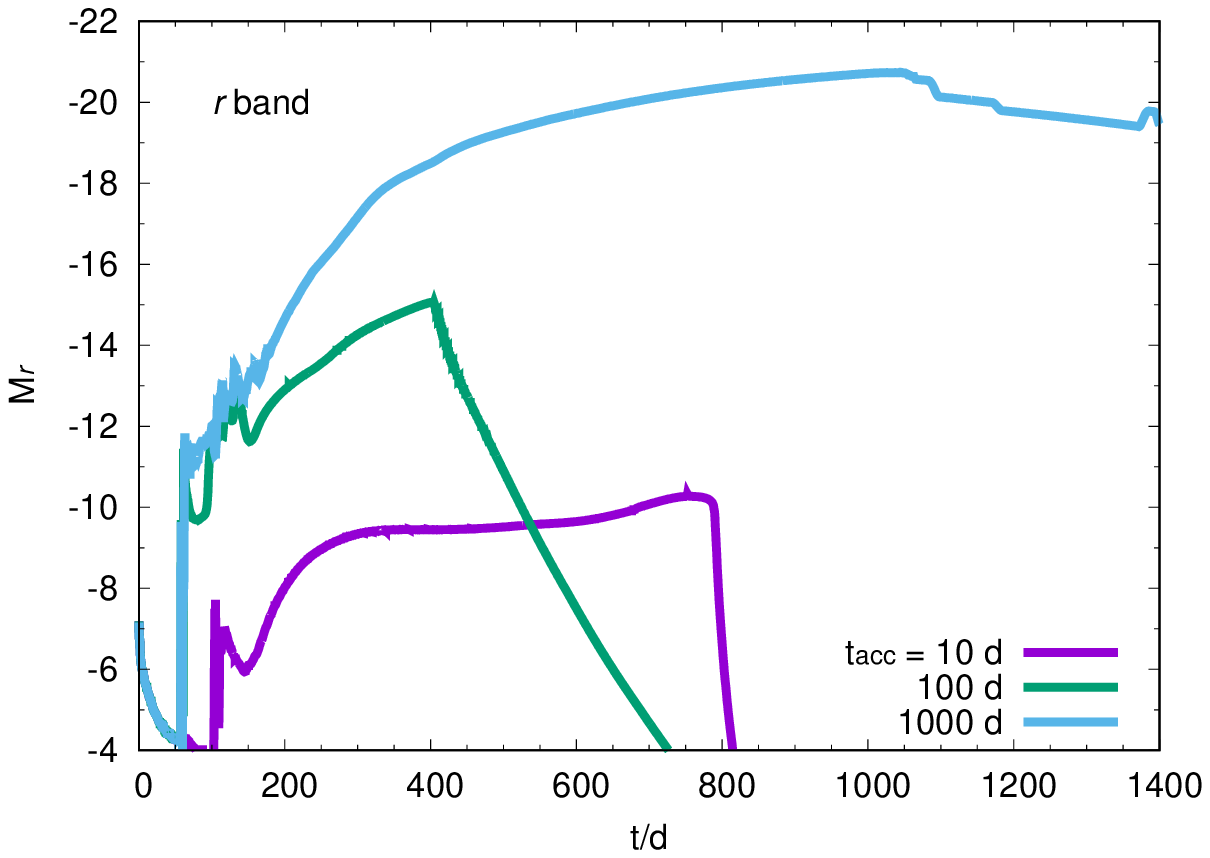}
	\includegraphics[width=\columnwidth]{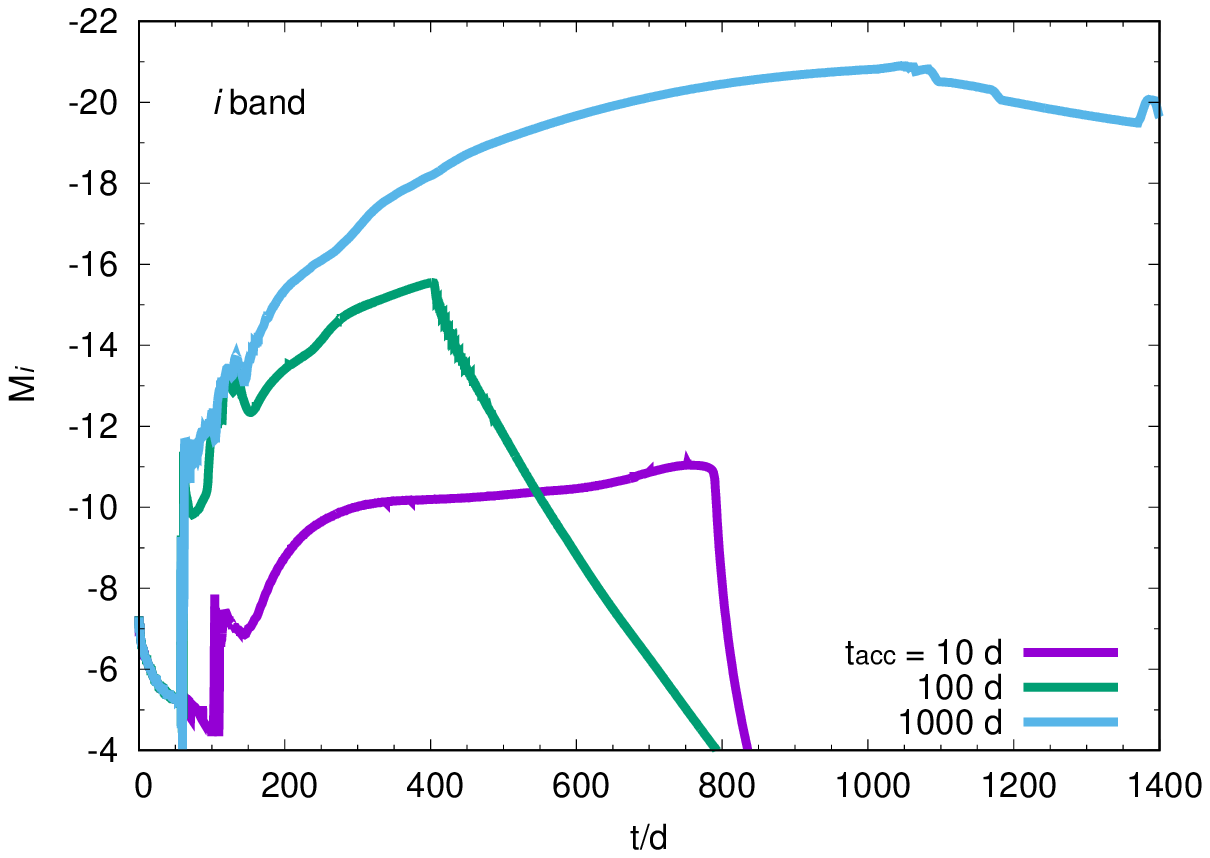}	
    \caption{Synthetic light curves of the \tzo\ explosions in the \textit{g}, \textit{r}, and \textit{i} bands. The time $t$ is after the onset of the \tzo\ collapse.}
    \label{fig:griband}
\end{figure}

Fig.~\ref{fig:bolometric} shows the bolometric light curves of the \tzo\ explosions with $t_\mathrm{acc}=10,100$ and 1000~d. Their photospheric temperature and velocity are presented in Fig.~\ref{fig:photosphere}. The synthetic \textit{g}, \textit{r} and \textit{i} band light curves are presented in Fig.~\ref{fig:griband}. Before the forward shock reaches the surface of the progenitor, the bolometric luminosity stays at the \tzo\ luminosity of around $5\times 10^{38}~\mathrm{erg~s^{-1}}$. The forward shock reaches the surface at 105~d ($t_\mathrm{acc}=10~\mathrm{d}$) and 51~d ($t_\mathrm{acc}=100$ and 1000~d). The models with $t_\mathrm{acc}=100$ and 1000~d have the same shock-appearance date because the energy deposition is the same in the two models at the beginning. 

The explosion with $t_\mathrm{acc}=10~\mathrm{d}$ is observed as a Type~IIP SN-like transient with a very long plateau duration (Figs~\ref{fig:bolometric} and \ref{fig:griband}), as analytically predicted by \citet{moriya2018}. The plateau duration is about 500~d and its bolometric luminosity is around $6\times 10^{39}~\mathrm{erg~s^{-1}}$, although the luminosity continues to increase slowly until the end of the plateau. The absolute magnitudes in the \textit{g}, \textit{r} and \textit{i} bands are around $-7.8$, $-9.5$, and $-10.3$, respectively, but they slightly depend on time. The photospheric velocity is relatively small (50 to $70~\mathrm{km~s^{-1}}$, Fig.~\ref{fig:photosphere}). The plateau phase is caused by the hydrogen recombination as in the case of Type~IIP SNe. Once the recombination wave reaches the bottom of the envelope the plateau phase ends and the luminosity drops.

The explosion model with $t_\mathrm{acc}=100~\mathrm{d}$ has a light curve with a steady luminosity increase from around 150~d (Figs.~\ref{fig:bolometric} and \ref{fig:griband}). The bolometric luminosity reaches a peak at 400~d with $3\times 10^{41}~\mathrm{erg~s^{-1}}$. The luminosity is sustained even after the termination of the energy injection at 100~d because of the hydrogen recombination as in the case of Type~IIP SNe. In this model, the recession of the recombination wave in the Lagrangian frame is slower than the ejecta expansion in the Eulerian frame. Thus, the luminosity increases as the recombination wave recedes towards the centre of the expanding envelope. The time of the luminosity peak corresponds to the moment when the recombination wave reaches at the bottom of the ejecta. Right after that the luminosity suddenly declines.
The photospheric velocity is around $500~\mathrm{km~s^{-1}}$ (Fig.~\ref{fig:photosphere}).

When the accretion energy input is sustained for 1000~d ($t_\mathrm{acc}=1000~\mathrm{d}$), the luminosity increases for 1000~d as seen in Figs~\ref{fig:bolometric} and \ref{fig:griband}. The luminosity peak is reached at 1000~d when the energy injection is terminated. The peak bolometric luminosity is $8\times 10^{43}~\mathrm{erg~s^{-1}}$. We find the peak optical magnitudes are around $-20.7$. The long-lasting luminosity source is the steady accretion towards the centre. The photospheric velocity is 1000 to $2000~\mathrm{km~s^{-1}}$ (Fig.~\ref{fig:photosphere}).

\section{Discussion}\label{sec:discussion}
In the previous section, we have found that \tzo\ outbursts should be observed as long-lasting transients with durations of more than 100~d. The expected luminosity range is diverse, ranging from those that are similar to massive stars (about $10^{39}~\mathrm{erg~s^{-1}}$) to SNe (about $10^{43}~\mathrm{erg~s^{-1}}$). The faintest \tzo\ explosions with $t_\mathrm{acc}=10~\mathrm{d}$ are difficult to discover because of their subtle luminosity change as well as their faintness. They could be discovered as vanishing stars after a short luminosity increase lasting for a few years. Such an event could be discovered during a survey for disappearing stars \citep[][]{kochanek2008,adams2017a,adams2017b,neustadt2021}.

\begin{figure}
	\includegraphics[width=\columnwidth]{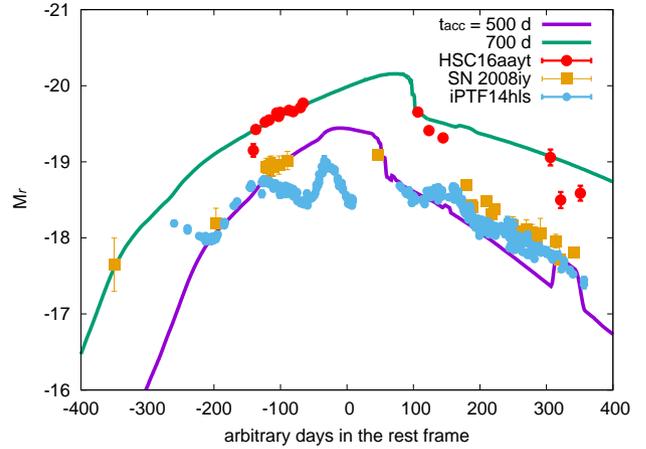}
    \caption{
    Comparison of synthetic \tzo\ light curves having $t_\mathrm{acc}=500$ and 700~d with long-lasting SNe, HSC16aayt \citep{moriya2019aayt}, SN~2008iy \citep{miller2010sn2008iy}, and iPTF14hls \citep{arcavi2017iptf14hls,sollerman2019iptf14hls}. Time is arbitrary shifted to match the observations and models.
    }
    \label{fig:comparison}
\end{figure}

If the energy input from the accretion is sustained long enough, the \tzo\ explosions become as bright as SNe and we expect to find them among SN candidates. We predict that \tzo\ explosions have much longer time-scales (more than $100~\mathrm{d}$) than typical SNe (less than $100~\mathrm{d}$). Thus, \tzo\ explosions can be discovered among unusually long-lasting SNe. In Fig.~\ref{fig:comparison}, we compare our \tzo\ light curve models with some long-lasting SNe. HSC16aayt \citep[][]{moriya2019aayt} and SN~2008iy \citep[][]{miller2010sn2008iy} are Type~IIn SNe showing narrow emission lines that are interpreted to originate from the interaction between dense circumstellar media and SN ejecta. The photospheric velocities of the \tzo\ explosions are predicted to be lower than those of SNe and so relatively narrow spectral features may actually originate from small photospheric velocities. It is also possible that the extended \tzo\ progenitors experience mass loss that leads to the narrow spectral features. The long-lasting SNe with the rise times of more than 100~d are promising candidates for \tzo\ explosions.

Another well-known mysterious long-lasting SN with an extremely long duration is iPTF14hls \citep{arcavi2017iptf14hls,sollerman2019iptf14hls}. Its origin is still not clear \citep[e.g.,][]{woosley2018,moriya2020}. While the duration is consistent with our \tzo\ explosion models, iPTF14hls has a photospheric velocity of $4000~\mathrm{km~s^{-1}}$ which is much higher than what we predict for our \tzo\ explosions. In addition, we do not expect to have the bumpy light curves as found for iPTF14hls (Fig.~\ref{fig:comparison}). Thus, iPTF14hls is not likely related to a \tzo\ explosions.

With an estimated Galactic \tzo\ birth rate of about $10^{-4}~\mathrm{yr^{-1}}$ \citep{podsiadlowski1995tzofate,ablimit2021} and a Galactic SN rate of about $10^{-2}~\mathrm{yr^{-1}}$ \citep{li2011snrate}, about 1\% of SNe may come from \tzo\ explosions\footnote{We note that the event rate estimate of \citet{moriya2018} has typos.}. SNe lasting for more than 100~d are rare and their event rates are not likely as high as 1\% of SNe. This may mean that most \tzo\ explosions are faint and difficult to discover. If the energy injection from the accretion disk is not usually sustained for a long time, most \tzo\ explosions would be observed only as faint transients. Thus, searching for disappearing stars that accompany faint transients is more likely a promising way to discover exploding \tzos. Other possibilities are that the \tzo\ birth rate is lower than $10^{-4}~\mathrm{yr^{-1}}$ or \tzos\ do not explode at all.

Distinguishing \tzo\ explosions from similar explosions that are triggered by iron core collapse is quite challenging. Although core collapse of massive stars around 16~\Msun\ usually ends up with successful SN explosions, it is possible that some of them fail to explode \citep[][]{ugliano2012,sukhbold2016}. If they form an accretion disk in a similar way to that of the \tzo\ collapse, we can have a similar transient. The identification of the progenitors is important to check if the explosions are of \tzos\ or not. 

We have so far approximated that the energy injection from the accretion disk is spherically symmetric. It is possible, however, that the accretion disk launches collimated jets. In such cases, the jets may lead to gamma-ray bursts with a very long duration as proposed for Swift 1644+57 \citep[][]{quataert2012kasen}, although several different progenitors may lead to such gamma-ray bursts with a very long duration \citep[e.g.,][]{soker2021}.

\section{Summary}\label{sec:summary}
We here investigated the observational properties of \tzo\ explosions. We found that \tzo\ explosions lead to transients with the durations of 100 to $1000~\mathrm{d}$. The possible luminosity range is quite diverse $10^{39}$ to $10^{44}~\mathrm{erg~s^{-1}}$. This depends on the duration of the energy input from the accretion disk triggering the \tzo\ explosions. The faintest \tzo\ explosions, which appear when the accretion energy injection is not sustained for long, have similar luminosities to those of massive stars. As in the case of Type~IIP SNe, the light curves are expected to have a long plateau phase. The plateau phase of \tzo\ explosions lasts for a few years. Their photospheric velocity is 50 to $70~\mathrm{km~s^{-1}}$. They may be observed as vanishing stars after brightening for several years. The bright \tzo\ explosions, which appear when the accretion energy injection lasts a long time, can be observed as SNe with rise times of several hundred days. We found that the expected light curves are similar to those of some long-lasting SNe such as HSC16aayt and SN~2008iy, although they are Type~IIn SNe. The photospheric velocity is expected to be of the order of $100~\mathrm{km~s^{-1}}$.

While we only have a few SNe that have long-duration light curves similar to our synthetic light curves from \tzo\ explosions, many long-duration SNe will be found in the coming era of the Rubin Observatory's Legacy Survey of Space and Time (LSST). Since the faintest \tzo\ explosions are expected to have luminosities that are similar to massive stars, they can be eventually found as disappeared stars.

\section*{Acknowledgements}
We thank the referee, Christopher Tout, for providing a constructive report that improved this manuscript.
T.J.M. thanks Emily Levesque for organizing the workshop ``An Exploration of Thorne-\.Zytkow Objects'' during the 238th meeting of the American Astronomical Society which initiated this work.
T.J.M. is supported by the Grants-in-Aid for Scientific Research of the Japan Society for the Promotion of Science (JP18K13585, JP20H00174, JP21K13966, JP21H04997).
S.I.B. is supported by grant RSF 19-12-00229 for the development of STELLA code.
This research has been supported in part by the RFBR (19-52-50014)-JSPS bilateral program.
Numerical computations were in part carried out on PC cluster at Center for Computational Astrophysics (CfCA), National Astronomical Observatory of Japan.


\section*{Data Availability}
The data underlying this article will be shared on reasonable request to the corresponding author.
 



\bibliographystyle{mnras}
\bibliography{mnras} 








\bsp	
\label{lastpage}
\end{document}